%% file: main.tex
\documentclass{easychair}

\usepackage{xcolor}

\usepackage{doc}

\usepackage{cite}

\usepackage{booktabs}
\usepackage{longtable}
\usepackage{pgfplotstable}
\usepackage{titlesec}
\usepackage{multirow}
\usepackage{subcaption}
\usepackage{fancyvrb}
\usepackage{cleveref}

\pgfplotsset{compat=1.16}

\pgfkeysifdefined{/pgfplots/table/output empty row/.@cmd}{
    \pgfplotstableset{
        empty header/.style={
          every head row/.style={output empty row},
        }
    }
}{
    \pgfplotstableset{
        empty header/.style={
            typeset cell/.append code={%
                \ifnum\pgfplotstablerow=-1 %
                    \pgfkeyssetvalue{/pgfplots/table/@cell content}{}%
                \fi
            }
        }
    }
}

\titlespacing*{\paragraph}
{0pt}{2pt}{5pt}

\title{The Second International Verification of Neural Networks Competition (VNN-COMP 2021): Summary and Results%
}

\author{Stanley Bak\thanks{S. Bak is with Stony Brooks, \tt stanley.bak@stonybrook.edu.}, Changliu Liu\thanks{C. Liu is with Carnegie Mellon University, \tt cliu6@andrew.cmu.edu.}, Taylor Johnson\thanks{T. Johnson is with Vanderbilt University, \tt taylor.johnson@vanderbilt.edu.}
}

\institute{
 }

\authorrunning{}
\date{\phantom{date}\\}
\titlerunning{VNN-COMP 2021 Report}

\begin{document}

\maketitle

\begin{abstract}
This report summarizes the second International Verification of Neural Networks Competition (VNN-COMP 2021), held as a part of the 4th Workshop on Formal Methods for ML-Enabled Autonomous Systems that was collocated with the 33rd International Conference on Computer-Aided Verification (CAV). Twelve teams participated in this competition.
The goal of the competition is to provide an objective comparison of the state-of-the-art methods in neural network verification, in terms of scalability and speed.
Along this line, we used standard formats (ONNX for neural networks and VNNLIB for specifications), standard hardware (all tools are run by the organizers on AWS), and tool parameters provided by the tool authors.
This report summarizes the rules, benchmarks, participating tools, results, and lessons learned from this competition.
\end{abstract}


\input{introduction}

\input{rules}

\input{participants}

\input{benchmarks}

\input{results}

\input{conclusion}

\input{acks}

\label{sect:bib}

\bibliographystyle{plain}
\bibliography{bib/nnv,bib/nnenum,bib/peregriNN,bib/verinet,bib/oval,bib/venus,bib/MIPVerify, bib/ERAN, bib/alpha-beta-CROWN,bib/debona,bib/dnnf,bib/nvjl,bib/nn4sys,bib/Marabou,bib/RPM}



\end{document}

%% file: introduction.tex
\section{Introduction}
\label{sec:introduction}

Methods based on machine learning are increasingly being deployed for a wide range of problems, including recommendation systems, machine vision and autonomous driving. While machine learning has made significant contributions to such applications, few tools provide formal guarantees about the behaviours of neural networks. 

In particular, for data-driven methods to be usable in safety-critical applications, including autonomous systems, robotics, cybersecurity, and cyber-physical systems, it is essential that the behaviours generated by neural networks are well-understood and can be predicted at design time. In the case of systems that are learning at run-time it is desirable that any change to the underlying system respects a given safety-envelope for the system.

While the literature on verification of traditionally designed systems is wide and successful, there has been a lack of results and efforts in this area until recently. The International Verification of Neural Networks Competition (VNN-COMP) was established in 2020, aiming to bring together researchers working on techniques for the verification of neural networks. In 2021, VNN-COMP\footnote{\url{https://sites.google.com/view/vnn2021/home}} was held as a part of the 4th Workshop on Formal Methods for ML-Enabled Autonomous Systems (FoMLAS) that was collocated with the 33rd International Conference on Computer-Aided Verification (CAV). 

\begin{figure}
    \centering
    \includegraphics[width=\linewidth]{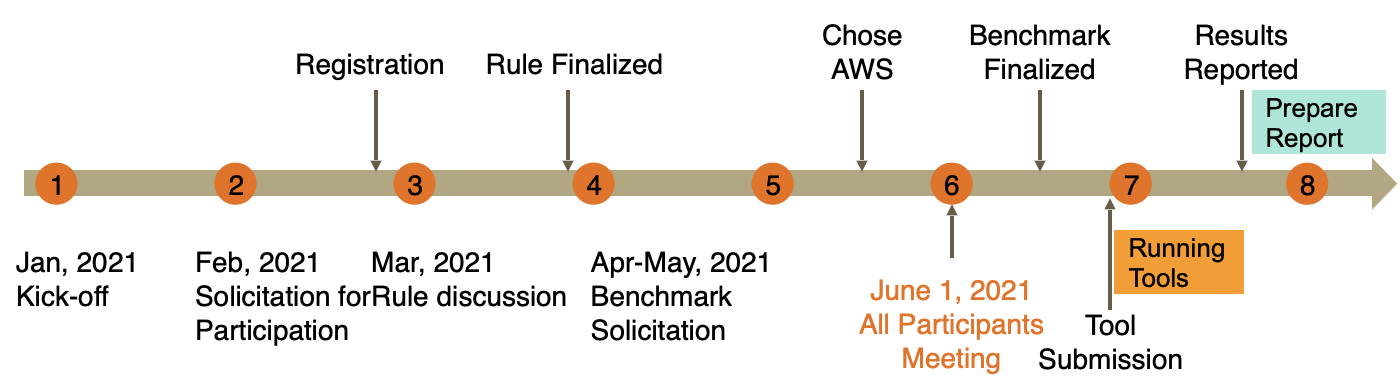}
    \caption{Timeline for VNN-COMP2021.}
    \label{fig:timeline}
\end{figure}

While the first VNN-COMP in 2020 was a friendly competition where the participants tested their tools and reported the results in parallel, the second VNN-COMP in 2021 aims to provide a fair comparison and standardize the pipeline of the competition. Such standardization includes 1) standard formats where we use ONNX for neural networks and VNNLIB for specifications; and 2) standard hardware where all tools are run by the organizers on AWS either on CPU instances or cost-equivalent GPU instances. The competition was kicked off in January and the solicitation for participation was sent in February 2021. By March, several teams registered to participate in this competition. The rule discussion was finalized in March 2021 where the finalized rules are summarized in \cref{sec:rules}. From April to May 2021, the benchmarks to test the tools were solicited. Meanwhile, after comparing different choices, the organizing team finally decided to use AWS as our testing platform. On June 1st, 2021, all participants had an online meeting to agree on the rules and benchmarks. By the end of June 2021, twelve teams submitted their tools and the organizers spent two weeks running the tools on AWS to obtain the final results. The final results were reported in FoMLAS on July 19, 2021. The timeline is summarized in \cref{fig:timeline}. Most discussions took place on the repository \url{https://github.com/stanleybak/vnncomp2021}. There are three issues: rules discussion, benchmarks discussion, and tool submission, serving as the venues for the corresponding discussions. Moreover, the repository hosts all the submitted benchmarks and the scripts to run the tests. 

The remainder of this report is organized as follows. \Cref{sec:rules} discusses the competition rules. \Cref{sec:participants} lists all participating tools, \Cref{sec:benchmarks} lists all benchmarks, and \Cref{sec:results} summarizes the results. \Cref{sec:conclusion} concludes the report and discusses potential future improvements.

%% file: rules.tex
\section{Rules}
\label{sec:rules}

\paragraph*{Terminology}
An \emph{instance} is defined as (specification (pre- and post-condition), network, timeout). 
For example: an MNIST classifier with one input image, a given local robustness threshold $\epsilon$, and a specific timeout.
A \emph{benchmark} is defined as a set of instances.
For example: a specific MNIST classifier with 100 input images, a given robustness threshold $\epsilon$, and one timeout per input.

\paragraph*{Run-time caps}
Per instance: any verification instance will timeout after at most X minutes, determined by the benchmark proposer. These can be different for each instance. 
Per benchmark: there is an upper runtime limit of 6 hours per benchmark. 
For example, a benchmark proposal could have six instances with a one hour timeout, or 100 instances each with a 3.6 minute timeout.
To provide a fair comparison, we quantify the startup overhead for each tool by running it on small networks; and then we subtract the overhead from the total runtime.

\paragraph*{Instance score}
Each instance is scored is as follows: 
\begin{itemize}\setlength{\itemsep}{0pt}
    \item Correct hold: 10 points;
    \item Correct violated (where random tests or simple adversarial example generation did not succeed): 10 points;
    \item Correct violated (where random tests or simple adversarial example generation succeeded): 1 point;
    \item Incorrect result: -100 points.
\end{itemize}

\paragraph*{Time bonus}
Time bonus is computed as follows.
\begin{itemize}\setlength{\itemsep}{0pt}
    \item The fastest tool for each solved instance will receive +2 points. 
    \item The second fastest tool will receive +1 point. 
\end{itemize}

All runtimes below 1.0 seconds after overhead correction (explained below) are considered to be 1.0 seconds exactly for scoring purposes. If two tools have runtimes within 0.2 seconds (after all corrections), for scoring purposes we will consider them the same runtime. 

\paragraph{Overhead Correction}
According to the rules discussion, we decided to subtract tool overhead time from the results. For example, simply importing tensorflow from Python and acquiring the GPU can sometimes take about 5 seconds, which would be unfortunate for benchmarks like ACASXu where some verification times are under a second.

To subtract overhead, we created trivial network instances and included those in the measurements. We then observed the minimum verification time along all instances, and considered that to be the overhead time for the tool.

One issue with that was that some tools had different overhead depending on if they were run in CPU mode or GPU mode, and this type of measurement penalized the GPU mode unintentionally. In the score reporting, we include a multi-overhead result where we apply the overhead measured for the mode the tool was actually run in.

\paragraph{Benchmark score} 
The benchmark score of each category is a percentage. 
It is computed as 100 times the sum of the individual instance scores for that benchmark category divided by the maximum sum of instance scores of any tool for that benchmark category. 
For example, the tool with the highest sum of instance scores for a category should get 100\%. 

\paragraph{Format}
This year we standardized the inputs to be \texttt{onnx} neural networks and \texttt{vnnlib} specification files.
Tool authors were also required to provide scripts to install their tool as much as possible, as well as run their tool on a specific instance provided the network file, specification file, and timeout.
Specifications included simple disjunctions in both pre- and post-conditions to encode properties like an unchanged classification for an input in one of multiple hyper-boxes.
The specification is regarded as encoding a counter-example, meaning that a property is proven ``correct'' if the specification is shown to be unsatisfiable, while the property is shown to be violated, if a counterexample fulfilling the specification is found. 
Hence, robustness with respect to inputs in a hyper-box has to be encoded as disjunctive property, where any of the other classes is constrained to become the maximum output.

%% file: participants.tex
\section{Participants}
\label{sec:participants}
The following tools and teams participated in VNN-COMP. They are summarized in \cref{tab:tool summary}.

\begin{table}[]
    \centering
    \begin{tabular}{c|cccc}
    \toprule
        Tool & 
        GPU? & 
        Floating Point Accuracy & 
        Use of External Solvers  \\
\midrule
Marabou &
- &
LP-default &
Gurobi \\
VeriNet &
- &
Either &
Xpress Solver
 \\
ERAN & 
Yes &
Sound &
Gurobi\\
$\alpha$,$\beta$-CROWN &
Yes &
Either &
Gurobi
 \\
DNNF &
- &
Either &
None
 \\
NNV &
- &
64bit &
Matlab
 \\
OVAL & 
Yes &
Either &
None \\
RPM &
- &
64bit &
None
 \\
NV.jl &
- &
64bit &
None
 \\
Venus &
- &
64bit&
Gurobi\\
Debona &
- &
32bit &
Gurobi \\
nnenum &
- &
Either & 
None
 \\ \bottomrule
    \end{tabular}
    \caption{Summary of key features of participating tools. For the ``GPU'' column, we only mark the tools that were tested on GPU during this competition. Tools that support GPU but were tested on CPU during this competition are not marked. For the floating point accuracy, ``either'' means the tool can handle either 32 or 64 bit floating point, depending on the onnx network; ``LP-default'' means the tool uses the default settings for LP solver; ``sound'' means the tool is fully floating-point-sound arithmetic with respect to IEEE-754 semantics up to the LP solver (up to 64x slower than non-fp-sound 32 bit arithmetic).}
    \label{tab:tool summary}
\end{table}


\subsection{Marabou}

\paragraph{Team} Guy Amir$^1$, Clark Barrett$^2$, Ahmed Irfan$^3$, Guy Katz$^1$, Teruhiro Tagomori$^4$, Alex Usvyatsov$^1$, Haoze Wu$^2$, Alexsandar Zeljic$^2$.

$^1$ Hebrew University of Jerusalem, $^2$ Stanford University, $^3$ Amazon Web Services, $^4$ NRI Secure.

\paragraph{Description} Marabou~\cite{katz2019marabou} is a Neural Network Verification toolkit that can answer queries about a network’s properties by encoding and solving these queries as constraint satisfaction problems. It can accommodate networks with different activation functions (including ReLU, Leaky ReLU, Sign \cite{amir2020smt}, Max, Absolute value) and topologies (e.g., FFNNs, CNNs, residual connections). It also uses the Split-and-Conquer algorithm~\cite{wu2020parallelization} for parallelization to further enhance scalability. Marabou accepts multiple input formats, including protocol buffer files generated by the popular TensorFlow framework for neural networks, and the ONNX format.

The core of Marabou resolves around the Reluplex procedure~\cite{katz2017reluplex}, but it also supports multiple new techniques and solving modes.
In particular, it incorporated the DeepPoly analysis introduced in \cite{DeepPoly:19} and the (MI)LP-based bound tightening first seen in \cite{Tjeng2019EvaluatingRO}. In terms of complete verification procedure, in addition to the Reluplex procedure, Marabou also supports solving the verification query with a MILP encoding.

For the competition, Marabou uses the DeepPoly analysis to tighten the variable bounds and then uses a portfolio strategy for complete verification, with a fraction of the CPUs solving a MILP-encoding of the verification problem with Gurobi, and the rest running a new complete verification procedure that is currently under submission.

\paragraph{Link} https://github.com/NeuralNetworkVerification/Marabou

\paragraph{Commit} For reproduction of VNN-Comp results, use \url{https://github.com/anwu1219/Marabou_private/commit/81e9f14f7ae9f6a2097524ea1291e86434ef42dc}.

\paragraph{Hardware and licenses} CPU, Gurobi License.

\paragraph{Participated benchmarks} \texttt{ACASXu}, \texttt{cifar10\_resnet},  \texttt{eran}, \texttt{marabou-cifar10}, \texttt{mnistfc}, \texttt{oval21}, \texttt{verivital}.

\subsection{VeriNet}

\paragraph{Team} 
Patrick Henriksen, Alessio Lomuscio (Imperial College London).

\paragraph{Description} VeriNet~\cite{HenriksenLomuscio20, HenriksenLomuscio21} is a complete Symbolic Interval Propagation (SIP) based verification toolkit for feed-forward neural networks. The underlying algorithm utilises SIP to create a linear abstraction of the network, which, in turn, is used in an LP-encoding to analyse the verification problem. A branch and bound-based refinement phase is used to achieve completeness. 

VeriNet implements various optimisations, including a gradient-based local search for counterexamples, optimal relaxations for Sigmoids, adaptive node splitting~\cite{HenriksenLomuscio20}, succinct LP-encodings, and a novel splitting heuristic that takes into account indirect effects splits have on succeeding relaxations~\cite{HenriksenLomuscio21}.

VeriNet supports a wide range of layers and activation functions, including Relu, Sigmoid, Tanh, fully connected, convolutional, max and average pooling, batch normalisation, reshape, crop and transpose operations, as well as additive residual connections.

Note that VeriNet subsumes the Deepsplit method presented in \cite{HenriksenLomuscio21}.

\paragraph{Link} Scheduled for release late August 2021 at: https://vas.doc.ic.ac.uk/software/neural/. 

\paragraph*{Hardware and licences}
CPU and GPU, Xpress Solver license required for large networks. 

\paragraph*{Participated benchmarks} \texttt{ACASXu}, \texttt{cifar10\_resnet}, \texttt{cifar2020}, \texttt{eran}, \texttt{marabou-cifar10}, \texttt{mnistfc}, \texttt{nn4sys}, \texttt{oval21}, \texttt{verivital}.

\subsection{ERAN}

\paragraph*{Team}
Mark Niklas Müller (ETH Zurich), Gagandeep Singh (UIUC), Markus Püschel (ETH Zurich), Martin Vechev (ETH Zurich)

\paragraph*{Description}
ERAN~\cite{SinghNIPS:18, DeepPoly:19, singh2019refinement, singh2019krelu, mueller2021prima, gpupoly} is a neural network verifier based on leveraging abstract interpretations to encode a network, pre- and post-condition as an LP or MILP problem. ERAN supports both incomplete and complete verification and can handle fully-connected, convolutional, and residual network architectures containing ReLU, Sigmoid, Tanh, and Maxpool non-linearities. It uses 64-bit precision (up to 16 times slower than 32-bit on some GPUs) and an arithmetic which is floating-point-sound  (performs 4 times more operations) with respect to  IEEE-754 semantics up to the LP solver. Single- and multi-neuron relaxations of non-linear activations~\cite{mueller2021prima} computed using GPUPoly \cite{gpupoly} are combined with a partial MILP encoding and neuron-wise bound-refinement \cite{singh2019refinement} to obtain a precise network encoding.  The analyzer is written in Python, uses ELINA~\cite{Singh:17} for numerical abstractions, and Gurobi for solving LP and MILP instances. While the use of both GPU and CPU enables ERAN to utilize all resources of a system it also requires both a relatively strong GPU and CPU to be available in order to avoid one bottlenecking the other (which occurred with the GPU AWS instance).

When run in complete mode, ERAN generates concrete counterexamples. In incomplete mode, ERAN attempts to falsify a property by running a PGD attack before attempting verification using increasingly more expensive and precise abstractions. As we use the same abstractions for all instances of a benchmark, ERAN might fail to verify a property before exceeding the timeout. In these cases, a more expensive abstraction might have been able to verify the property in time. 

\paragraph*{Link}
\url{https://github.com/eth-sri/eran}

\paragraph*{Commit}
Please use the main repository for anything but reproducing VNN-COMP results.
\url{https://github.com/mnmueller/eran_vnncomp2021.git}\\ (1e474c77f72f86f450df9f0a860b4d35c490ea7c)

\paragraph*{Hardware and licences}
CPU and GPU, GUROBI License

\paragraph*{Participated benchmarks} \texttt{ACASXu}, \texttt{cifar10\_resnet}, \texttt{cifar2020}, \texttt{eran}, \texttt{marabou-cifar10}, \texttt{mnistfc}, \texttt{nn4sys}, \texttt{oval21}, \texttt{verivital}.

\subsection{$\alpha,\!\beta$-CROWN}

\paragraph*{Team}
Huan Zhang\textsuperscript{*} (Carnegie Mellon), Kaidi Xu\textsuperscript{*} (Northeastern), Shiqi Wang\textsuperscript{*} (Columbia), Zhouxing Shi (UCLA), Yihan Wang (UCLA), Xue Lin (Northeastern), Suman Jana (Columbia), Cho-Jui Hsieh (UCLA), Zico Kolter (Carnegie Mellon); * indicates equal contribution.
\paragraph*{Description} The $\alpha,\!\beta$-CROWN (\texttt{alpha-beta-CROWN}) verifier is based on an efficient bound propagation algorithm, CROWN~\cite{zhang2018efficient}, with a few crucial extensions~\cite{xu2020automatic,xu2021fast,wang2021betacrown}. We use the generalized version of CROWN in the \texttt{auto\_LiRPA} library~\cite{xu2020automatic} which supports general neural network architectures (including convolutional layers, residual connections, recurrent neural networks and Tranformers) and a wide range of activation functions (e.g., ReLU, tanh, sigmoid,  max pooling and average pooling), and is efficiently implemented on GPUs. We jointly optimize intermediate layer bounds and final layer bounds using gradient ascent (referred to as $\alpha$-CROWN or optimized CROWN/LiRPA~\cite{xu2021fast}). Additionally, we use branch and bound~\cite{bunelunified2018} (BaB) and incorporate split constraints in BaB into the bound propagation procedure efficiently via the $\beta$-CROWN algorithm~\cite{wang2021betacrown}. The combination of efficient, optimizable and GPU accelerated bound propagation with BaB produces a powerful and scalable neural network verifier.

Our verifier also utilizes a mixed integer programming (MIP) solver (Gurobi) for networks where MIP runs relatively fast, following the formulation in~\cite{Tjeng2019EvaluatingRO}. We use MIP to solve the tightest possible intermediate layer bounds for as many neurons as possible on CPUs within the timeout budget, and use $\alpha$-CROWN to solve the remaining ones on GPUs. Finally, we conduct BaB with $\beta$-CROWN using tightened bounds. Although the GPU AWS instance has weak CPUs, we still find that MIP is helpful for some benchmarks, and it can become more beneficial on a machine with both strong CPUs and GPUs. Note that $\alpha$-CROWN can exceed the power of a typical LP verifier when intermediate layer bounds are jointly  optimized~\cite{xu2021fast,salman2019convex}, so we do not use a LP solver to tighten bounds.
\paragraph*{Link} \url{https://github.com/huanzhang12/alpha-beta-CROWN}
\paragraph*{Commit} \texttt{c12e6eeaf6b16f99a99b65f377d0f450d6466a83} (only for reproducing competition results; please use the \texttt{main} branch version for other proposes)
\paragraph*{Hardware and licenses} CPU and GPU with 32-bit or 64-bit floating point; Gurobi license required for \texttt{mnistfc}, \texttt{eran}, \texttt{marabou-cifar10} and \texttt{verivital} benchmarks.
\paragraph*{Participated benchmarks} \texttt{ACASXu}, \texttt{cifar10\_resnet}, \texttt{cifar2020}, \texttt{eran}, \texttt{marabou-cifar10}, \texttt{mnistfc}, \texttt{nn4sys}, \texttt{oval21}, \texttt{verivital}.

\subsection{DNNF}

\paragraph{Team} 
David Shriver, Sebastian Elbaum, Matt Dwyer (University of Virginia).

\paragraph{Description} DNNF~\cite{shriver-etal:ICSE:2021:dnnf} is a tool for neural network property falsification. It only attempts to find counter-examples to a property specification, and will not prove that a property holds. DNNF reduces properties and networks to robustness problems which can then be falsified using many different adversarial attack methods. Our reduction approach enables us to support much more complex property and network specifications, such as specifications with direct input and output relations, such as the relation $y > x$. The backend falsification method used for VNN-COMP is a custom implementation of PGD, but DNNF can also be run with several off-the-shelf methods from foolbox or cleverhans, two popular python packages for adversarial attacks.

DNNF makes use of the DNNV framework~\cite{shriver-etal:CAV:2021:dnnv} to load networks and properties, as well as to perform network simplifications.
The current implementation of DNNF supports many different network operations. In particular it supports ONNX models with the following operations: Add, Atan, AveragePool, BatchNormalization, Concat, Conv, ConvTranspose, Elu, Flatten, Gather, Gemm, LeakyRelu, MatMul, MaxPool, Mul, Relu, Reshape, Resize, Shape, Sigmoid, Softmax, Sub, Tanh, Transpose, Unsqueeze.

DNNF can also be run on GPUs, which can speed up falsification for large models.

\paragraph*{Link} \url{https://github.com/dlshriver/DNNF}

\paragraph*{Commit} VNN-COMP results can be reproduced with commit\\
\texttt{d4f08b43e4ad622157c65ac071183a3a0f4e6fe0}. For other uses, we suggest the \texttt{main} branch.

\paragraph*{Hardware and licences}
DNNF can be run on the CPU or GPU, with no additional licenses required.

\paragraph*{Participated benchmarks} \texttt{ACASXu}, \texttt{cifar10\_resnet}, \texttt{cifar2020}, \texttt{eran}, \texttt{marabou-cifar10}, \texttt{mnistfc}, \texttt{nn4sys}, \texttt{oval21}, \texttt{verivital}.

\subsection{NNV}
\paragraph*{Team} Neelanjana Pal (Vanderbilt University), Taylor T Johnson (Vanderbilt University)
\paragraph*{Description}

The Neural Network Verification Tool (NNV) \cite{xiang2018tnnls,tran2019formalise,tran2019fm,tran2020cav,tran2020cav_tool} is written primarily with Matlab and implements reachability-analysis methods for neural network verification with a particular focus on applications of closed-loop neural network control systems in autonomous cyber-physical systems. NNV uses geometric representations such as star sets that allows for a layer-by-layer computation of the exact reachable set for feed-forward deep neural networks. In the event that a particular safety property is violated, NNV can construct and visualize the complete set of counterexample inputs for a neural network.

\paragraph{Link} https://github.com/verivital/nnv.
\paragraph*{Commit} \texttt{3ca2629aaceb9080e4d08a0f9c6b51854f9c7b7f} (for reproducing competition results; otherwise please use the master version).

\paragraph*{Hardware and licences} GPU and CPU with 64-bit floating point. A license for Matlab will be required.
\paragraph*{Participated benchmarks} \texttt{ACASXu}, \texttt{cifar2020}, \texttt{eran}, \texttt{mnistfc}, \texttt{oval21}, \texttt{verivital}.

\subsection{OVAL}
\paragraph*{Team} Alessandro De Palma (University of Oxford), Florian Jaeckle (University of Oxford), M. Pawan Kumar (University of Oxford)
\paragraph*{Description}

The OVAL verification tool is an optimization-based complete verifier that relies on a specialized Branch and Bound (BaB) framework for neural network verification. 
In this context, a BaB method is composed of three main components (see \cite{bunel2020branch} for an overview): a \emph{branching strategy} to divide the verification property into easier subproblems, a \emph{bounding algorithm} to compute over-approximation bounds for each subproblem, and a \emph{falsification algorithm} to look for counter-examples to the property. \vspace{5pt}

For networks with medium to large input dimensionality, OVAL relies on the efficient FSB~\cite{depalma2021improved} branching strategy, which combines a dual-based scoring of ReLU neurons~\cite{bunel2020branch} with inexpensive strong branching approximations to select an activation to split upon. For networks of small input dimensionality, OVAL can revert to input splitting as in \cite{bunelunified2018}. 
Bounds on the subproblems are obtained by adaptively choosing~\cite{depalma2021sparsealgos} between bounding algorithms of varying degrees of tightness: the competition entry relies on Beta-CROWN~\cite{wang2021betacrown}, which effectively solves the convex hull of element-wise activations, and Active Set~\cite{depalma2021scaling}, which operates on the tighter relaxation from \cite{Anderson2020} to tackle harder properties.
In addition, the framework supports a variety of bounding algorithms \cite{bunel2020lagrangian,depalma2021sparsealgos,Dvijotham2018,xu2021fast}.
The search for counter-examples is performed using the MI-FGSM~\cite{dong2018boosting} adversarial attack, which we adapted to perform general property falsification. \vspace{5pt}
 
The implementation of the OVAL framework, written in PyTorch~\cite{Paszke2017}, exploits GPU acceleration and is massively parallel over both the BaB subproblems and the relative intermediate computations \cite{bunel2020lagrangian}. It currently supports fully connected and convolutional networks, with ReLU, maxpool and average pooling layers.

\paragraph{Link} https://github.com/oval-group/oval-bab.
\paragraph*{Commit} \texttt{014b6ee5071508430c8e515bbae725306db68fe1} in order to reproduce competition results. We otherwise suggest to employ the master version.
\paragraph*{Hardware and licences} GPU and CPU with 32-bit or 64-bit floating point. No license is required.
\paragraph*{Participated benchmarks} \texttt{ACASXu}, \texttt{cifar2020}, \texttt{eran}, \texttt{marabou-cifar10}, \texttt{mnistfc}, \texttt{nn4sys}, \texttt{oval21}, \texttt{verivital}.

\subsection{RPM}

\paragraph*{Team}
Joe Vincent (Stanford), Mac Schwager (Stanford)

\paragraph*{Description}
The Reachable Polyhedral Marching (RPM) tool \cite{vincent2021reachable} is a method for computing exact forward and backward reachable sets of feedforward neural networks with ReLU activation. Verification problems are posed as backward reachability problems. A unique feature of the RPM tool is its incremental computation of reachable sets. For verification this means that unsafe inputs may be found before computing the complete reachable set, leading to early termination. RPM does not currently have a parallel implementation, although the algorithm is amenable to parallelization.

\paragraph*{Link}
\url{https://github.com/StanfordMSL/Neural-Network-Reach/tree/vnn_comp_2021}
\paragraph*{Commit} \texttt{861ce6e380e3cc2d439a7bca87b59817e4624af6} for reproducing competition results. For other purposes the most recent commit is suggested.

\paragraph*{Hardware and licences}
CPU, no license is required.

\paragraph*{Participated benchmarks}
\texttt{ACASXu}

\subsection{ComposableNeuralVerification (NV.jl)}

\paragraph*{Team}
Tianhao Wei (Carnegie Mellon), Chen Tan (Northeastern), Changliu Liu (Carnegie Mellon)

\paragraph*{Description}
This tool is adapted from the original NeuralVerification.jl~\cite{liu2019neuralverification} developed at the Stanford Intelligent System Lab. This Julia toolbox implemented a wide variety of verification algorithms that use reachability, optimization, and search, which are summarized in~\cite{liu2021algorithms}. We added support for onnx format networks and vnnlib format specifications,

\paragraph*{Link}
\url{https://github.com/intelligent-control-lab/NeuralVerification.jl}

\paragraph*{Hardware and licences}
CPU, no licence

\paragraph{Commit} 4e612602ba4b34b42416742d85476d9b0dcdcb51 (for reproducing competition results; otherwise please use the master branch)

\paragraph*{Participated benchmarks}
nn4sys and AcasXu

\subsection{Venus}

\paragraph*{Team}
Panagiotis Kouvaros (Imperial College London), Alessio Lomuscio (Imperial College London)

\paragraph*{Description}
Venus is a complete verification tool for Relu-based feed-forward neural
networks.  Venus implements a MILP-based verification method whereby it
leverages dependency relations between the ReLU nodes to reduce the search
space that needs to be considered during branch-and-bound. The dependency
relations are exploited via callback cuts~\cite{Botoeva+20} and via a branching method that
divides the verification problem into a set of sub-problems whose MILP
formulations require fewer integrality constraints~\cite{KouvarosLomuscio21a}.  To derive tight MILP
encodings, Venus additionally implements a symbolic interval propagation method
for computing the pre-activation bounds of the ReLU nodes; the method optimises
the linear relaxation of each of the ReLU nodes towards minimising the
over-approximation error in subsequent layers.

\paragraph*{Link} \url{https://github.com/vas-group-imperial/
venus2}

\paragraph*{Commit} For the reproduction of the VNN-COMP2021 results please use the repository
\url{https://github.com/pkouvaros/venus2_vnncomp21} (57e9608041d230b5d78c4f2afb890b81035436a1).

\paragraph*{Hardware and licenses} CPU, GUROBI License.

\paragraph*{Participated benchmarks} ACASXU, mnistfc, nn4sys.

\subsection{Debona}

\paragraph*{Team}
Christopher Brix (RWTH Aachen University), Thomas Noll (RWTH Aachen University)

\paragraph*{Description}
Debona is a fork of VeriNet \cite{HenriksenLomuscio20}. However, the abstract domain used by VeriNet defines symbolic linear upper and lower bounds that are parallel to each other, i.e., offset only by some scalar value. On the contrary, Debona utilizes independent upper and lower bounds. This allows for a tighter relaxation especially for ReLU operations, where a lower bound of zero may be better than bounds that are negative in large regions of the input space. This idea has been described in \cite{brix2020debona} but was independently previously published in \cite{DeepPoly:19}.

\paragraph*{Link}
\url{https://github.com/ChristopherBrix/Debona}

\paragraph*{Commit}
f000f3d483b2cc592233d0ba2a1a0327210562c8

\paragraph*{Hardware and licences}
CPU, Gurobi licence

\paragraph*{Participated benchmarks}
AcasXu, eran, mnistfc and nn4sys

\subsection{nnenum}
\paragraph*{Team} Stanley Bak (Stony Brook Univeristy)

\paragraph*{Description} 
The nnenum tool uses multiple levels of abstraction to achieve high-performance verification of ReLU networks without sacrificing completeness~\cite{bak2020vnn}. Analysis combines three types of zonotopes with star set (triangle) overapproximations~\cite{tran2019fm}, and uses efficient parallelized ReLU case splitting~\cite{bak2020cav}. 
The ImageStar method~\cite{tran2020cav} allows sets to be quickly propagated through all layers supported by the ONNX runtime, such as convolutional layers with arbitrary parameters.  
The tool is written in Python 3, uses GLPK for LP solving.
New this year we added support for \texttt{vnnlib} files, and optimized some of the LP timeout parameters for acasxu~\cite{bak2021nnenum}.

\paragraph*{Link}
\url{https://github.com/stanleybak/nnenum}

\paragraph*{Commit}
c93a39cb568f58a26015bd151acafab34d2d4929

\paragraph*{Hardware and licences}
CPU, No licenses required

\paragraph*{Participated benchmarks}
AcasXu, cifar2020, mnistfc, oval

%% file: benchmarks.tex
\section{Benchmarks}
\label{sec:benchmarks}


\begin{table}[]
    \centering
    \begin{tabular}{c|ccc}
    \toprule
         Benchmark &
Application &
Network Types &
Largest NN 
\\
\midrule
Acasxu &
Control &
Feedforward + ReLU Only &
54.6k 
\\
Cifar10\_resnet &
Image Classification &
ResNet &
487k 
\\
Cifar2020 (unscored) &
Image Classification &
Conv + ReLU &
9.41M 
\\
Eran &
Image Classification &
Feedforward + non-ReLU &
1.68M 
\\
Marabou-cifar10 &
Image Classification &
Conv + ReLU &
1.29M 
\\
Mnistfc &
Image Classification &
Feedforward + ReLU Only &
2.03M 
\\
nn4sys &
Database Indexing &
Feedforward + ReLU Only &
336.5M* 
\\
Oval21 &
Image Classification &
Conv + ReLU &
840k 
\\
Verivital &
Image Classification &
Conv + maxpool / avgpool &
46.3k 
\\
\bottomrule

    \end{tabular}
    \footnotesize *{After zipping, the network is of 1.79M.}
    \caption{Overview of all benchmarks. }
    \label{tab:my_label}
\end{table}

\subsection{ACAS Xu}
\paragraph{Networks} The ACASXu benchmarks consists of ten properties defined over 45 neural networks used to issue turn advisories to aircraft to avoid collisions. The neural networks have 300 neurons arranged in 6 layers, with ReLU activation functions. There are five inputs corresponding to the aircraft states, and five network outputs, where the minimum output is used as the turn advisory the system ultimately produces.

\paragraph{Specifications} We use the original 10 properties~\cite{katz2017reluplex}, where properties 1-4 are checked on all 45 networks as was done in later work by the original authors~\cite{katz2019marabou}. Properties 5-10 are checked on a single network. The total number of benchmarks is therefore 186. The original verification times ranged from seconds to days---including some benchmark instances that did not finish. This year we used a timeout of around two minutes (116 seconds) for each property, in order to fit within a total maximum runtime of six hours.

\subsection{Cifar10\_resnet}

\paragraph*{Proposed by} the $\alpha,\!\beta$-CROWN team.

\paragraph*{Motivations} Currently, many tools are hard-coded to handle feedforward networks only. To make neural network verification more useful in practical scenarios, we advocate that tools should handle more general architecture. Residual networks~\cite{he2016deep} (ResNet) is the first step towards this goal due to its relatively simple structure and practical significance. The propose of this benchmark is to provide some incentives for the community to develop more generic tools.

\paragraph*{Networks} We provided two ResNet models on CIFAR-10 image classification task with the following structures:
\begin{itemize}
    \item \texttt{ResNet-2B} with 2 residual blocks: 5 convolutional layers + 2 linear layers
    \item \texttt{ResNet-4B} with 4 residual blocks: 9 convolutional layers + 2 linear layers
\end{itemize}
The networks are trained via adversarial training with an $\ell_\infty$ perturbation norm of $\epsilon=\frac{2}{255}$. For simplicity, these networks do not contain batch normalization or pooling layers, and use ReLU activation functions. The ResNet-4B model is relatively large with over 10K neurons.

We evaluated both networks using a 100-step projected gradient descent (PGD) attack with 5 random restarts, and a simple bound propagation based verification algorithm CROWN~\cite{zhang2018efficient} (mathematically equivalent to the abstract interpretation used in DeepPoly~\cite{DeepPoly:19}) under $\ell_\infty$ norm perturbations. The results are listed in Table~\ref{tab:resnet_model}.

\begin{table}[htb]
\begin{tabular}{ccccccc}
\toprule
\multirow{2}{*}{Model} & \multirow{2}{*}{\# ReLUs} & \multirow{2}{*}{Clean acc.} & \multicolumn{2}{c}{$\epsilon=2/255$} & \multicolumn{2}{c}{$\epsilon=1/255$} \\
                       &                           &                             & PGD acc.    & Verified acc.   & PGD acc.    & Verified acc.   \\ \midrule
\texttt{ResNet-2B}              & 6244                      & 69.25\%                     & 54.82\%     & 26.88\%         & 62.24\%     & 57.16\%         \\ \hline
\texttt{ResNet-4B}              & 14436                     & 77.20\%                     & 61.41\%     & 0.24\%         & 69.75\%      & 23.28\%         \\ \bottomrule
\end{tabular}
\caption{Clean accuracy, PGD accuracy and CROWN verified accuracy for ResNet models. Note that the verified accuracy is obtained via the vanilla version of CROWN/DeepPoly which has been widely used as a simple baseline, not the $\alpha,\!\beta$-CROWN tool used in the competition.}
\label{tab:resnet_model}
\end{table}

To ensure the appropriate level of difficulty, we use $\epsilon=\frac{2}{255}$ for the ResNet-2B model and $\epsilon=\frac{1}{255}$ for the ResNet-4B model.

\paragraph*{Specifications} We randomly select 48 images from the CIFAR-10 test set for the ResNet-2B model and 24 images for the ResNet-4B model. The images are classified correctly and cannot be attacked by a 100-step PGD attack with 5 random restarts. For each image, we specify the property that the logit of the ground-truth label is always greater than the logits of all other 9 labels within $\ell_\infty$ norm input perturbation of $\epsilon=\frac{2}{255}$ for ResNet-2B and $\epsilon=\frac{1}{255}$ for ResNet-4B. The per-example timeout is set to 5 minutes and the overall runtime is guaranteed to be less than 6 hours.

\subsection{Cifar2020 (unscored)}

\paragraph*{Motivation} This benchmark combines two convolutional CIFAR10 networks from last year's VNN-COMP 2020 with a new, larger network with the goal to evaluate the progress made by the whole field of Neural Network verification.

\paragraph*{Networks} The two ReLU networks \texttt{cifar\_10\_2\_255} and \texttt{cifar\_10\_8\_255} with two convolutional and two fully-connected layers were trained for $\ell_\infty$ perturbations of $\epsilon = \frac{2}{255}$ and $\frac{8}{255}$, respectively, using COLT \cite{balunovic:20}  and the larger \texttt{ConvBig} with four convolutional and three fully-connected networks, was trained using adversarial training \cite{madry:17} and $\epsilon = \frac{2}{255}$.

\paragraph*{Specifications} We draw the first 100 images from the CIFAR10 test set and for every network reject incorrectly classified ones. For the remaining images, the specifications describe a correct classification under an $\ell_\infty$-norm perturbation of at most $\frac{2}{255}$ and $\frac{8}{255}$ for \texttt{cifar\_10\_2\_255} and \texttt{ConvBig} and \texttt{cifar\_10\_8\_255}, respectively and allow a per sample timeout of 5 minutes.

\subsection{eran}

\paragraph*{Proposed by:} The ERAN team

\paragraph*{Motivation} While most Neural Network Verificaiton methods focus their analysis on ReLU based networks, many modern network architectures, e.g., EfficientNet \cite{tan2019efficientnet}, are based on non-piecwise-linear activation functions. To begin to understand how the choice of activation function affects certifiability, the eran benchmark aims at comparing the certifiabilty of networks based on piecewise-linear and non-piecwise-linear activation functions under an $\ell_\infty$-norm based adversary.

\paragraph*{Networks} We consider a ReLU network with 8 hidden layers of width 200 and a Sigmoid network with 6 hidden layers of width 200. Both networks were trained using standard training.

\paragraph*{Specifications} We sample random images from the MNIST test set until we obtain 36 correctly classified images per network. For these images, the specifications describe a correct classification under an $\ell_\infty$-norm perturbation of at most 0.015 and 0.012 for the ReLU and Sigmoid network, respectively, and allow a per sample timeout of 5 minutes.

\subsection{Marabou-cifar10}

\paragraph{Proposed by} The Marabou team.

\paragraph*{Networks} This benchmark contains three convolutional networks, \texttt{cifar10\_small.onnx},
\texttt{cifar10\_medium.onnx}, and \texttt{cifar10\_large.onnx}, trained on the CIFAR10 dataset. Each network has 2 convolutional layers followed by two fully connected feed-forward layers. Each layer uses the ReLU activation functions. The networks are all trained with Adam optimizer for 120 epochs with learning rate 0.0002. The three networks contain 2568, 4944, and 10528 ReLUs, respectively. The test accuracy are $63.14\%$, $70.21\%$, and $74.16\%$, respectively. 
The networks expect the input image to be normalized between 0 and 1.

\paragraph*{Specifications} We randomly sample correctly classified images from the CIFAR10 test set. The specifications are targeted adversarial robustness, which states that the network does not mis-classify an image as a given adversarial label under $l_\infty$-norm perturbations. The target label is chosen as $(correctLabel + 1) \mod 10$. We propose two perturbation bounds: 0.012 and 0.024, and allow a per-query timeout of 5 minutes.

\subsection{Mnistfc}

\paragraph{Proposed by} The VeriNet team.

\paragraph*{Motivation} This benchmark contains fully connected networks with ReLU activation functions and varying depths. 

\paragraph*{Networks} The benchmark set consists of three fully-connected classification networks with 2, 4 and 6 
layers and 256 ReLU nodes in each layer trained on the MNIST dataset. The networks were first presented in a benchmark in VNN-COMP 2020~\cite{vnn-comp}.

\paragraph*{Specifications} We randomly sampled 15 correctly classified images from the MNIST test set. For each network and image, the specification was a correct classification under $l_\infty$ perturbations of at most $\epsilon = 0.03$ and $\epsilon = 0.05$. The timeouts were 2 minutes per instance for the 2-layer network and 5 minutes for the remaining two networks. 

\subsection{NN4Sys}
\paragraph*{Proposed by} The ComposableNeuralverification team
\paragraph*{Application}
The benchmark contains networks for database indexing, which is a 1D to 1D mapping. 

\begin{itemize}

\item \textit{Background}: learned index is a neural network (NN) based database index proposed by Kraska et al.~\cite{kraska18case}, 2018.
It shows great potential but has one drawback---for non-existing keys (i.e., the keys that do not exist in the database),
the outputs of a learned index can be arbitrary.

\item \textit{What we do}: to provide safety guarantees for \textit{all} keys,
we design a specification to dictate how ``far" one predicted position can be,
compared to its actual position (or the positions that non-existing keys should be).

\item \textit{What to verify}: our benchmark provides multiple pairs of (1) learned indexes (trained NNs)
and (2) corresponding specifications. We design these pairs with different parameters such
that they cover a variety of user needs and have varied difficulties for verifiers. 

\item \textit{Translating learned indexes to a VNN benchmark}: 
the original learned index~\cite{kraska18case} contains two stages (of NNs) for high precision.
However, this cascading structure is inconvenient/unsupported to verify
because there is a ``switch" operation---choosing one NN in the second stage
based on the prediction of the first stage's NN.
To convert learned indexes to a standard form, we merge the NNs in both stages
into an integrated network by adding some hand crafted layers.

\item \textit{A note on broader impact}: using NNs for systems is a broad topic, but many existing works
lack strict safety guarantees. We believe that NN Verification can help system developers gain confidence
to apply NNs to critical systems. We hope our benchmark can be an early step towards this vision.

\end{itemize}

\paragraph*{Networks}
The networks are feedforward with ReLU activations, and they are sparse networks.
There are six networks in this benchmark.
Three of them have original size 194.2M and zipped size 1.79M;
the other three have original size 336.5M and zipped size 790k.
This is because onnx does not support directly encoding of sparse matrices,
hence the networks are stored as fully connected networks.

\paragraph*{Specifications}
The specification aims to check if the prediction error is bounded.
The specification is a collection of pairs of input and output intervals such that
any input in the input interval should be mapped to the corresponding output interval.

\subsection{Oval21}


\paragraph*{Proposed by} The OVAL team.

\paragraph*{Motivations}

The majority of adversarial robustness benchmarks  consider image-independent perturbation radii, possibly resulting in some properties that are either easily verified by all verification methods, or too hard to be verified (for commonly employed timeouts) by any of them.
In line with the OVAL verification dataset from VNN-COMP 2020~\cite{vnn-comp}, whose versions have already been used in various recent works~\cite{bunel2020lagrangian,Lu2020Neural,depalma2021improved,depalma2021scaling,depalma2021sparsealgos,wang2021betacrown,xu2021fast,jaeckle2021generating, jaeckle2021neural}, the OVAL21 benchmark associates to each image-network pair a perturbation radius found via binary search to ensure that all properties are challenging to solve.

\paragraph*{Networks}
The benchmark includes 3 ReLU-based convolutional networks which were robustly trained~\cite{Wong2018} against $\ell_{\infty}$ perturbations of radius $\epsilon=2/255$ on CIFAR10.
Two of the networks, named \texttt{base} and \texttt{wide}, are composed of 2 convolutional layers followed by 2 fully connected layers and have respectively $3172$ and $6244$ activations. The third model, named \texttt{deep}, has 2 additional convolutional layers and a total of $6756$ activations.

\paragraph*{Specifications}
The verification properties represent untargeted adversarial
robustness (with respect to all possible misclassifications) to $\ell_{\infty}$ perturbations of varying $\epsilon$, with a per-instance timeout of $720$ seconds.
The property generation procedure relies on commonly employed lower and upper bounds to the adversarial loss to exclude perturbation radii that yield trivial properties.
10 correctly classified images per network are randomly sampled from the entire CIFAR10 test set, and a distinct $\epsilon \in [0, 16/255]$ is associated to each. First, a binary search is run to find the largest $\epsilon$ value for which a popular iterative adversarial attack \cite{dong2018boosting} fails to find an adversarial example. Then, a second binary search is run to find the smallest $\epsilon$ value for which bounds~\cite{xu2021fast} from the element-wise convex hull of the activations (with fixed intermediate bounds from \cite{Wong2018, zhang2018efficient}) fail to prove robustness. Both binary search procedures are run with a tolerance of $\epsilon_{\text{tol}}=0.1$. Denoting $\epsilon_{lb}$ as the smallest output from the two routines, and $\epsilon_{ub}$ as the largest, the following perturbation radius is chosen: $\epsilon = \frac{1}{3} \epsilon_{lb} + \frac{2}{3}\epsilon_{ub}$.

\paragraph*{Link} \url{https://github.com/stanleybak/vnncomp2021/tree/main/benchmarks/oval21}

\subsection{Verivital}
\paragraph{Proposed by} The VeriVITAL team.

\paragraph*{Motivation} Neural networks with pooling layers are vastly used in several applications. The main motivation for proposing this benchmark was to include the pooling layers as part of this year's VNN-Comp.

\paragraph*{Networks} This benchmark contains two MNIST classifiers with pooling layers, one with averagepooling layers and the other with maxpooling.

\paragraph*{Specifications} We randomly sampled 20 correctly classified images from the MNIST test set. For the network with averagepooling layers, the specification was to correctly classify those randomly chosen images with an $l_\infty$ perturbation radii($\epsilon$) of $0.02$ and $0.04$ and a timeout of 5 minutes. For the network with maxpooling layers the corresponding radius was $0.004$ with a timeout of 7 minutes.

\paragraph*{Link} \url{https://github.com/stanleybak/vnncomp2021/tree/main/benchmarks/verivital}

%% file: results.tex
\newpage
\section{Results}
\label{sec:results}

Each tool was run on all benchmarks which produced a \texttt{csv} file of results.
This was sent to the authors for review, which sometimes required rerunning certain benchmarks to make sure they match the expected results.
Python code was then written to process the results and compute the scores.
The final \texttt{csv} files for each tool as well as scoring scripts are available online: \url{https://github.com/stanleybak/vnncomp2021_results}.

This also includes detailed log files for each benchmark showing the specific runtime for each tool and the score awarded.
This can be used to find challenging instances to help with tool development.
For example, in the output files in the repo you may see things like:
\begin{Verbatim}[frame=single]
Row: ['ACASXU_run2a_4_2_batch_2000-prop_2', '-', '6.4 (h)', '10.5 (h)', 
      'timeout', '41.1 (h)', 'timeout', 'timeout', '64.8 (h)', '62.5 (h)', 
      'timeout', 'timeout', 'timeout', '-']
73: nnv score: 0
73: nnenum score: 12
73: venus2 score: 11
73: NN-R score: 0
73: VeriNet score: 10
73: DNNF score: 0
73: Debona score: 0
73: a-b-CROWN score: 10
73: oval score: 10
73: Marabou score: 0
73: ERAN score: 0
73: NV.jl score: 0
73: randgen score: 0
\end{Verbatim}

The tools are listed in order, and row is the times and result for each tool. So \texttt{nnenum} should be holds with a time of 6.4 (after subtracting overhead). 
The scores are also listed for each tool. Since \texttt{nnenum} was the fastest on this instance, it got 12 points, 10 for correct plus 2 for time bonus as fastest. 
The second fastest was \texttt{venus2} at 10.5 seconds, so they get 11 points. None of the remaining tools were within 0.2 seconds, so they all received 10 points if they completed analysis successfully.

\subsection{Overall Score}

The overall score for VNN-COMP 2021 is shown in Table~\ref{tab:score}.
Two tables are included, based on the two ways to measure overhead discussed in Section~\ref{sec:rules}.
Overall, $\alpha,\!\beta$-CROWN performed best, followed by VeriNet.
We awarded two third place results, one to oval and one to ERAN, as their ranking depended upon factors like overhead as well as how incorrect results were judged, discussed next.
For all the remaining tables in this section, the numbers reported correspond to the multi-overhead measurements.

One unexpected aspect was how to judge incorrect results, since tools currently are not required to produce a counter-example when an instance is falsified.
We considered two reasonable options, which was voting (majority is assumed to be correct), and odd-one-out.
In odd-one-out, only if a single tool's output differs from all the others is the result is considered incorrect.
If multiple tools produce the same result or if only two tools completed the instance and their results differ, then the instance is ignored for scoring purposes.
This generally had a slight effect on the score, which was significant enough to affect the order in the total ranking.
Specifically, the positions of ERAN and oval for third place could be affected by the scoring parameters, when using the "Single Overhead" overhead correction.
Notice that, however, both voting and odd-one-out are imperfect ways to judge incorrect results, and it may be the case that the mismatching tool was in fact correct.
In future editions of VNN-COMP we may standardize counter-example outputs and require they are produced when instances can be falsified, to remedy this shortcoming.
For the reported category scores, we display odd-one-out scores and results.

Other statistics and the individual benchmark scores are also included below.
In general, the GPU tools did better, as well as tools that could support a large number of benchmarks and verified a large number of benchmark instances.
Several tools produced mismatching results, and an interesting followup study could identify the underlying reasons for this unsoundness.

\begin{table}[!htb]
    \caption{VNN-COMP 2021 Overall Score}
    \label{tab:score}
    \begin{subtable}{.5\linewidth}
      \centering
        \caption{Single Overhead}
{\setlength{\tabcolsep}{2pt}
\begin{tabular}[h]{@{}llr@{}}
\toprule
\textbf{\# ~} & \textbf{Tool} & \textbf{Score}\\
\midrule
1 & $\alpha$,$\beta$-CROWN & 779.2 \\
2 & VeriNet & 701.2 \\
3 & oval & 582.0 \\
4 & ERAN & 581.1 \\
5 & Marabou & 335.3 \\
6 & Debona & 201.9 \\
7 & venus2 & 189.2 \\
8 & nnenum & 184.5 \\
9 & nnv & 57.2 \\
10 & NV.jl & 48.1 \\
11 & RPM & 25.4 \\
12 & DNNF & 24.3 \\
13 & randgen & 1.9 \\
\bottomrule
\end{tabular}
}
    \end{subtable}%
    \begin{subtable}{.5\linewidth}
      \centering
        \caption{Multi-Overhead}
        {\setlength{\tabcolsep}{2pt}
\begin{tabular}[h]{@{}llr@{}}
\toprule
\textbf{\# ~} & \textbf{Tool} & \textbf{Score}\\
\midrule
1 & $\alpha$,$\beta$-CROWN & 779.7 \\
2 & VeriNet & 705.0 \\
3 & ERAN & 643.4 \\
4 & oval & 581.8 \\
5 & Marabou & 339.0 \\
6 & Debona & 201.9 \\
7 & venus2 & 189.2 \\
8 & nnenum & 184.6 \\
9 & nnv & 57.2 \\
10 & NV.jl & 48.1 \\
11 & RPM & 25.4 \\
12 & DNNF & 24.3 \\
13 & randgen & 1.9 \\
\bottomrule
\end{tabular}
}
    \end{subtable} 
\end{table}

\clearpage
\subsection{Other Stats}

This section presents other statistics related to the measurements that are interesting, but did not play a direct role in scoring this year.
With ERAN, which had different overheads, the `CPU Mode' column in Table~\ref{tab:overhead} corresponds to the overhead used for the ACASXu and ERAN benchmarks, whereas the `Seconds' column corresponds to when the GPU was used (all others).

\begin{table}[h]
\begin{center}
\caption{Overhead} \label{tab:overhead}
{\setlength{\tabcolsep}{2pt}
\begin{tabular}[h]{@{}llrr@{}}
\toprule
\textbf{\# ~} & \textbf{Tool} & \textbf{Seconds} & \textbf{~~CPU Mode}\\
\midrule
1 & Marabou & 0.2 & - \\
2 & randgen & 0.3 & - \\
3 & nnenum & 1.0 & - \\
4 & venus2 & 1.7 & - \\
5 & DNNF & 2.0 & - \\
6 & VeriNet & 2.2 & - \\
7 & Debona & 2.5 & - \\
8 & oval & 5.1 & - \\
9 & $\alpha$,$\beta$-CROWN & 6.1 & - \\
10 & ERAN & 7.1 & 3.7 \\
11 & nnv & 8.4 & - \\
12 & NV.jl & 20.9 & - \\
13 & RPM & 52.2 & - \\
\bottomrule
\end{tabular}
}
\end{center}
\end{table}


\begin{table}[h]
\begin{center}
\caption{Num Benchmarks Participated} \label{tab:stats0}
{\setlength{\tabcolsep}{2pt}
\begin{tabular}[h]{@{}llr@{}}
\toprule
\textbf{\# ~} & \textbf{Tool} & \textbf{Count}\\
\midrule
1 & VeriNet & 9 \\
2 & ERAN & 9 \\
3 & $\alpha$,$\beta$-CROWN & 9 \\
4 & oval & 8 \\
5 & Marabou & 7 \\
6 & nnv & 6 \\
7 & DNNF & 5 \\
8 & randgen & 4 \\
9 & nnenum & 4 \\
10 & Debona & 4 \\
11 & venus2 & 3 \\
12 & NV.jl & 2 \\
13 & RPM & 1 \\
\bottomrule
\end{tabular}
}
\end{center}
\end{table}


\begin{table}[h]
\begin{center}
\caption{Num Instances Verified} \label{tab:stats1}
{\setlength{\tabcolsep}{2pt}
\begin{tabular}[h]{@{}llr@{}}
\toprule
\textbf{\# ~} & \textbf{Tool} & \textbf{Count}\\
\midrule
1 & $\alpha$,$\beta$-CROWN & 766 \\
2 & VeriNet & 717 \\
3 & ERAN & 656 \\
4 & oval & 636 \\
5 & Marabou & 364 \\
6 & nnenum & 310 \\
7 & Debona & 280 \\
8 & venus2 & 266 \\
9 & nnv & 141 \\
10 & NV.jl & 97 \\
11 & RPM & 72 \\
12 & DNNF & 55 \\
13 & randgen & 33 \\
\bottomrule
\end{tabular}
}
\end{center}
\end{table}


\begin{table}[h]
\begin{center}
\caption{Num Violated} \label{tab:stats2}
{\setlength{\tabcolsep}{2pt}
\begin{tabular}[h]{@{}llr@{}}
\toprule
\textbf{\# ~} & \textbf{Tool} & \textbf{Count}\\
\midrule
1 & ERAN & 177 \\
2 & $\alpha$,$\beta$-CROWN & 177 \\
3 & oval & 175 \\
4 & VeriNet & 175 \\
5 & Marabou & 103 \\
6 & nnenum & 73 \\
7 & Debona & 70 \\
8 & venus2 & 63 \\
9 & DNNF & 55 \\
10 & RPM & 44 \\
11 & randgen & 33 \\
12 & NV.jl & 11 \\
\bottomrule
\end{tabular}
}
\end{center}
\end{table}


\begin{table}[h]
\begin{center}
\caption{Num Holds} \label{tab:stats3}
{\setlength{\tabcolsep}{2pt}
\begin{tabular}[h]{@{}llr@{}}
\toprule
\textbf{\# ~} & \textbf{Tool} & \textbf{Count}\\
\midrule
1 & $\alpha$,$\beta$-CROWN & 589 \\
2 & VeriNet & 542 \\
3 & ERAN & 479 \\
4 & oval & 461 \\
5 & Marabou & 261 \\
6 & nnenum & 237 \\
7 & Debona & 210 \\
8 & venus2 & 203 \\
9 & nnv & 141 \\
10 & NV.jl & 86 \\
11 & RPM & 28 \\
\bottomrule
\end{tabular}
}
\end{center}
\end{table}


\begin{table}[h]
\begin{center}
\caption{Mismatched (Incorrect) Results} \label{tab:stats4}
{\setlength{\tabcolsep}{2pt}
\begin{tabular}[h]{@{}llr@{}}
\toprule
\textbf{\# ~} & \textbf{Tool} & \textbf{Count}\\
\midrule
1 & Marabou & 26 \\
2 & Debona & 14 \\
3 & nnv & 12 \\
4 & NV.jl & 5 \\
5 & venus2 & 1 \\
\bottomrule
\end{tabular}
}
\end{center}
\end{table}

\clearpage

\subsection{Benchmark Scores}
The results for the individual categories are shown below.
For overall score, the tools which participated in all or almost all of the benchmarks did best.
Within individual benchmarks, some tools performed well despite not ranking high in the overall score.
The results presented here are for multi-overhead setup, with incorrect results scored using the odd-one-out strategy.
Adjusting these parameters produced minor changes in the overall score and rankings, but was omitted for clarity.
If these alternate scores are of interest, the discussion at the beginning of Section~\ref{sec:results} outlines where to access the scripts used to compute scores.

\input{cats}

\clearpage

%% file: cats.tex

\begin{table}[h]
\begin{center}
\caption{Benchmark \texttt{acasxu}} \label{tab:cat_{cat}}
{\setlength{\tabcolsep}{2pt}
\begin{tabular}[h]{@{}lllllrr@{}}
\toprule
\textbf{\# ~} & \textbf{Tool} & \textbf{Verified} & \textbf{Falsified} & \textbf{Fastest} & \textbf{Score} & \textbf{Percent}\\
\midrule
1 & nnenum & 138 & 47 & 155 & 1910 & 100.0\% \\
2 & VeriNet & 138 & 47 & 117 & 1852 & 97.0\% \\
3 & Marabou & 137 & 46 & 115 & 1809 & 94.7\% \\
4 & oval & 138 & 47 & 98 & 1794 & 93.9\% \\
5 & venus2 & 138 & 46 & 94 & 1778 & 93.1\% \\
6 & $\alpha$,$\beta$-CROWN & 138 & 47 & 67 & 1732 & 90.7\% \\
7 & ERAN & 125 & 46 & 24 & 1506 & 78.8\% \\
8 & Debona & 84 & 42 & 39 & 1086 & 56.9\% \\
9 & RPM & 28 & 44 & 9 & 486 & 25.4\% \\
10 & nnv & 29 & 0 & 29 & 348 & 18.2\% \\
11 & DNNF & 0 & 41 & 12 & 182 & 9.5\% \\
12 & randgen & 0 & 28 & 0 & 28 & 1.5\% \\
13 & NV.jl & 45 & 9 & 0 & -23 & 0\% \\
\bottomrule
\end{tabular}
}
\end{center}
\end{table}


\begin{table}[h]
\begin{center}
\caption{Benchmark \texttt{cifar10-resnet}} \label{tab:cat_{cat}}
{\setlength{\tabcolsep}{2pt}
\begin{tabular}[h]{@{}lllllrr@{}}
\toprule
\textbf{\# ~} & \textbf{Tool} & \textbf{Verified} & \textbf{Falsified} & \textbf{Fastest} & \textbf{Score} & \textbf{Percent}\\
\midrule
1 & $\alpha$,$\beta$-CROWN & 58 & 0 & 12 & 623 & 100.0\% \\
2 & VeriNet & 48 & 0 & 29 & 548 & 88.0\% \\
3 & ERAN & 43 & 0 & 36 & 502 & 80.6\% \\
4 & Marabou & 39 & 0 & 0 & 390 & 62.6\% \\
\bottomrule
\end{tabular}
}
\end{center}
\end{table}


\begin{table}[h]
\begin{center}
\caption{Benchmark \texttt{cifar2020}} \label{tab:cat_{cat}}
{\setlength{\tabcolsep}{2pt}
\begin{tabular}[h]{@{}lllllrr@{}}
\toprule
\textbf{\# ~} & \textbf{Tool} & \textbf{Verified} & \textbf{Falsified} & \textbf{Fastest} & \textbf{Score} & \textbf{Percent}\\
\midrule
1 & oval & 146 & 41 & 174 & 2209 & 100.0\% \\
2 & $\alpha$,$\beta$-CROWN & 148 & 42 & 43 & 1996 & 90.4\% \\
3 & VeriNet & 139 & 42 & 5 & 1822 & 82.5\% \\
4 & ERAN & 107 & 43 & 132 & 1749 & 79.2\% \\
5 & nnenum & 62 & 13 & 0 & 741 & 33.5\% \\
6 & randgen & 0 & 2 & 0 & 2 & 0.1\% \\
7 & nnv & 6 & 0 & 0 & -140 & 0\% \\
\bottomrule
\end{tabular}
}
\end{center}
\end{table}


\begin{table}[h]
\begin{center}
\caption{Benchmark \texttt{eran}} \label{tab:cat_{cat}}
{\setlength{\tabcolsep}{2pt}
\begin{tabular}[h]{@{}lllllrr@{}}
\toprule
\textbf{\# ~} & \textbf{Tool} & \textbf{Verified} & \textbf{Falsified} & \textbf{Fastest} & \textbf{Score} & \textbf{Percent}\\
\midrule
1 & $\alpha$,$\beta$-CROWN & 60 & 1 & 26 & 670 & 100.0\% \\
2 & VeriNet & 48 & 1 & 49 & 588 & 87.8\% \\
3 & ERAN & 46 & 1 & 0 & 470 & 70.1\% \\
4 & Debona & 47 & 2 & 39 & 375 & 56.0\% \\
5 & oval & 21 & 0 & 18 & 247 & 36.9\% \\
6 & Marabou & 19 & 0 & 0 & 190 & 28.4\% \\
7 & nnv & 10 & 0 & 0 & 100 & 14.9\% \\
\bottomrule
\end{tabular}
}
\end{center}
\end{table}


\begin{table}[h]
\begin{center}
\caption{Benchmark \texttt{marabou-cifar10}} \label{tab:cat_{cat}}
{\setlength{\tabcolsep}{2pt}
\begin{tabular}[h]{@{}lllllrr@{}}
\toprule
\textbf{\# ~} & \textbf{Tool} & \textbf{Verified} & \textbf{Falsified} & \textbf{Fastest} & \textbf{Score} & \textbf{Percent}\\
\midrule
1 & $\alpha$,$\beta$-CROWN & 1 & 52 & 52 & 625 & 100.0\% \\
2 & ERAN & 0 & 52 & 51 & 613 & 98.1\% \\
3 & oval & 0 & 53 & 44 & 611 & 97.8\% \\
4 & VeriNet & 0 & 52 & 16 & 543 & 86.9\% \\
5 & Marabou & 0 & 29 & 0 & 281 & 45.0\% \\
6 & DNNF & 0 & 2 & 0 & 11 & 1.8\% \\
7 & randgen & 0 & 1 & 0 & 1 & 0.2\% \\
\bottomrule
\end{tabular}
}
\end{center}
\end{table}


\begin{table}[h]
\begin{center}
\caption{Benchmark \texttt{mnistfc}} \label{tab:cat_{cat}}
{\setlength{\tabcolsep}{2pt}
\begin{tabular}[h]{@{}lllllrr@{}}
\toprule
\textbf{\# ~} & \textbf{Tool} & \textbf{Verified} & \textbf{Falsified} & \textbf{Fastest} & \textbf{Score} & \textbf{Percent}\\
\midrule
1 & $\alpha$,$\beta$-CROWN & 49 & 21 & 35 & 772 & 100.0\% \\
2 & VeriNet & 39 & 21 & 57 & 716 & 92.7\% \\
3 & Debona & 37 & 22 & 47 & 688 & 89.1\% \\
4 & oval & 37 & 21 & 46 & 676 & 87.6\% \\
5 & ERAN & 34 & 22 & 47 & 654 & 84.7\% \\
6 & Marabou & 35 & 19 & 0 & 540 & 69.9\% \\
7 & venus2 & 31 & 16 & 26 & 522 & 67.6\% \\
8 & nnenum & 35 & 12 & 21 & 512 & 66.3\% \\
9 & nnv & 27 & 0 & 8 & 186 & 24.1\% \\
10 & DNNF & 0 & 7 & 0 & 70 & 9.1\% \\
\bottomrule
\end{tabular}
}
\end{center}
\end{table}


\begin{table}[h]
\begin{center}
\caption{Benchmark \texttt{nn4sys}} \label{tab:cat_{cat}}
{\setlength{\tabcolsep}{2pt}
\begin{tabular}[h]{@{}lllllrr@{}}
\toprule
\textbf{\# ~} & \textbf{Tool} & \textbf{Verified} & \textbf{Falsified} & \textbf{Fastest} & \textbf{Score} & \textbf{Percent}\\
\midrule
1 & $\alpha$,$\beta$-CROWN & 70 & 5 & 73 & 878 & 100.0\% \\
2 & VeriNet & 68 & 3 & 0 & 719 & 81.9\% \\
3 & ERAN & 67 & 4 & 0 & 704 & 80.2\% \\
4 & oval & 56 & 4 & 0 & 589 & 67.1\% \\
5 & NV.jl & 41 & 2 & 0 & 422 & 48.1\% \\
6 & venus2 & 34 & 1 & 0 & 250 & 28.5\% \\
7 & DNNF & 0 & 4 & 0 & 22 & 2.5\% \\
8 & randgen & 0 & 2 & 0 & 2 & 0.2\% \\
9 & Debona & 42 & 4 & 0 & -722 & 0\% \\
\bottomrule
\end{tabular}
}
\end{center}
\end{table}


\begin{table}[h]
\begin{center}
\caption{Benchmark \texttt{oval21}} \label{tab:cat_{cat}}
{\setlength{\tabcolsep}{2pt}
\begin{tabular}[h]{@{}lllllrr@{}}
\toprule
\textbf{\# ~} & \textbf{Tool} & \textbf{Verified} & \textbf{Falsified} & \textbf{Fastest} & \textbf{Score} & \textbf{Percent}\\
\midrule
1 & oval & 12 & 2 & 11 & 164 & 100.0\% \\
2 & $\alpha$,$\beta$-CROWN & 12 & 2 & 2 & 146 & 89.0\% \\
3 & VeriNet & 11 & 2 & 6 & 145 & 88.4\% \\
4 & ERAN & 6 & 2 & 3 & 86 & 52.4\% \\
5 & Marabou & 4 & 2 & 1 & 63 & 38.4\% \\
6 & nnenum & 2 & 1 & 0 & 30 & 18.3\% \\
7 & nnv & 16 & 0 & 4 & -31 & 0\% \\
\bottomrule
\end{tabular}
}
\end{center}
\end{table}


\begin{table}[h]
\begin{center}
\caption{Benchmark \texttt{verivital}} \label{tab:cat_{cat}}
{\setlength{\tabcolsep}{2pt}
\begin{tabular}[h]{@{}lllllrr@{}}
\toprule
\textbf{\# ~} & \textbf{Tool} & \textbf{Verified} & \textbf{Falsified} & \textbf{Fastest} & \textbf{Score} & \textbf{Percent}\\
\midrule
1 & $\alpha$,$\beta$-CROWN & 53 & 7 & 52 & 704 & 100.0\% \\
2 & oval & 51 & 7 & 57 & 694 & 98.6\% \\
3 & ERAN & 51 & 7 & 56 & 693 & 98.4\% \\
4 & VeriNet & 51 & 7 & 0 & 580 & 82.4\% \\
5 & DNNF & 0 & 1 & 0 & 10 & 1.4\% \\
6 & nnv & 53 & 0 & 0 & -168 & 0\% \\
7 & Marabou & 27 & 7 & 0 & -2260 & 0\% \\
\bottomrule
\end{tabular}
}
\end{center}
\end{table}

%% file: conclusion.tex
\section{Conclusion and Ideas for Future Competitions}
\label{sec:conclusion}
This report summarizes the 2$^\text{nd}$ Verification of Neural Networks Competition (VNN-COMP) held in 2021.
Improvements to the competition structure have been made, including standardization of common input formats (\texttt{onnx} and \texttt{vnnlib}), and common measurement hardware.
Based on the common benchmarks (CIFAR2020 and ACASXU), tools exhibited significant progress in terms of scalability and speed compared with previous years.
The comparison is imperfect, as last year we did not have standardized hardware, so it is unclear how much of the speed improvement is due to algorithmic improvements and how much is due to better hardware.
Future editions of the competition may better judge the year-to-year improvements in neural network verification methods.

The benchmarks and tool execution scripts are openly available for others to replicate: \url{https://github.com/stanleybak/vnncomp2021}.
We hope this serves as a fair comparison for evaluating future improvements to verification methods in upcoming publications.
From an applicability perspective, having a common input format hopefully reduces the barriers of industry participants to use the developed tools.

In future editions of VNN-COMP, some improvements we can make would be to allow more types of hardware. 
For example, any AWS EC2 machine could be chosen by the tool authors with roughly the same cost (\$3 an hour this year).
Multiple tool authors expressed that their tool could work faster if given both a strong CPU and GPU together.
Alternatively, we could allow custom hardware per benchmark.
This would likely require additional automation with the competition measurements, especially dealing with installing license files.
Some ideas for automating this could be to have authors provide the Gurobi licences to use, rather than the competition organizers.
Another improvement would be to improve overhead measurement, as detailed in Section~\ref{sec:rules}.
Finally, we should standardize the counter-example format, so that the ground truth for mismatched verification results can be provided.

%% file: acks.tex
\section*{Acknowledgements}
%

This competition (including the cost of AWS and the cash prizes) is supported by a gift from the Lu Jin Family Foundation. 

This research was supported in part by the Air Force Research Laboratory Information Directorate, through the Air Force Office of Scientific Research Summer Faculty Fellowship
Program, Contract Numbers FA8750-15-3-6003, FA9550-15-0001 and FA9550-20-F-0005.
This material is based upon work supported by the Air Force Office of Scientific Research under award numbers FA9550-19-1-0288 and FA9550-21-1-0121, the National Science Foundation (NSF) under grant number FMitF 1918450, and the Defense Advanced Research Projects Agency (DARPA) Assured Autonomy program through contract number FA8750-18-C-0089.
Any opinions, finding, and conclusions or recommendations expressed in this material are those of the author(s) and do not necessarily reflect the views of the United States Air Force, DARPA, nor NSF.

Tool authors listed in \cref{sec:participants} participated in the preparation and review of this report.

%% file: main.bbl
\begin{thebibliography}{10}

\bibitem{amir2020smt}
Guy Amir, Haoze Wu, Clark Barrett, and Guy Katz.
\newblock An smt-based approach for verifying binarized neural networks.
\newblock {\em arXiv preprint arXiv:2011.02948}, 2020.

\bibitem{Anderson2020}
Ross Anderson, Joey Huchette, Will Ma, Christian Tjandraatmadja, and Juan~Pablo
  Vielma.
\newblock Strong mixed-integer programming formulations for trained neural
  networks.
\newblock {\em Mathematical Programming}, 2020.

\bibitem{bak2020vnn}
Stanley Bak.
\newblock Execution-guided overapproximation (ego) for improving scalability of
  neural network verification, 2020.

\bibitem{bak2021nnenum}
Stanley Bak.
\newblock nnenum: Verification of relu neural networks with optimized
  abstraction refinement.
\newblock In {\em NASA Formal Methods Symposium}, pages 19--36. Springer, 2021.

\bibitem{bak2020cav}
Stanley Bak, Hoang-Dung Tran, Kerianne Hobbs, and Taylor~T. Johnson.
\newblock Improved geometric path enumeration for verifying {ReLU} neural
  networks.
\newblock In {\em 32nd International Conference on Computer-Aided Verification
  (CAV)}, July 2020.

\bibitem{Botoeva+20}
E.~Botoeva, P.~Kouvaros, J.~Kronqvist, A.~Lomuscio, and R.~Misener.
\newblock Efficient verification of neural networks via dependency analysis.
\newblock In {\em Proceedings of the 34th AAAI Conference on Artificial
  Intelligence (AAAI20)}. {AAAI} Press, 2020.

\bibitem{brix2020debona}
Christopher Brix and Thomas Noll.
\newblock Debona: Decoupled boundary network analysis for tighter bounds and
  faster adversarial robustness proofs.
\newblock {\em CoRR}, abs/2006.09040, 2020.

\bibitem{bunel2020lagrangian}
Rudy Bunel, Alessandro De~Palma, Alban Desmaison, Krishnamurthy Dvijotham,
  Pushmeet Kohli, Philip~HS Torr, and M~Pawan Kumar.
\newblock Lagrangian decomposition for neural network verification.
\newblock {\em Conference on Uncertainty in Artificial Intelligence}, 2020.

\bibitem{bunel2020branch}
Rudy Bunel, Jingyue Lu, Ilker Turkaslan, P~Kohli, P~Torr, and M~Pawan Kumar.
\newblock Branch and bound for piecewise linear neural network verification.
\newblock {\em Journal of Machine Learning Research}, 21(2020), 2020.

\bibitem{bunelunified2018}
Rudy Bunel, Ilker Turkaslan, Philip~HS Torr, Pushmeet Kohli, and M~Pawan Kumar.
\newblock A unified view of piecewise linear neural network verification.
\newblock {\em Advances in Neural Information Processing Systems}, 2018.

\bibitem{depalma2021scaling}
Alessandro De~Palma, Harkirat~Singh Behl, Rudy Bunel, Philip H.~S. Torr, and
  M.~Pawan Kumar.
\newblock Scaling the convex barrier with active sets.
\newblock In {\em International Conference on Learning Representations}, 2021.

\bibitem{depalma2021sparsealgos}
Alessandro De~Palma, Harkirat~Singh Behl, Rudy Bunel, Philip H.~S. Torr, and
  M.~Pawan Kumar.
\newblock Scaling the convex barrier with sparse dual algorithms.
\newblock {\em arXiv preprint arXiv:2101.05844}, 2021.

\bibitem{depalma2021improved}
Alessandro De~Palma, Rudy Bunel, Alban Desmaison, Krishnamurthy Dvijotham,
  Pushmeet Kohli, Philip~HS Torr, and M~Pawan Kumar.
\newblock Improved branch and bound for neural network verification via
  lagrangian decomposition.
\newblock {\em arXiv preprint arXiv:2104.06718}, 2021.

\bibitem{dong2018boosting}
Yinpeng Dong, Fangzhou Liao, Tianyu Pang, Hang Su, Jun Zhu, Xiaolin Hu, and
  Jianguo Li.
\newblock Boosting adversarial attacks with momentum.
\newblock In {\em Proceedings of the IEEE conference on computer vision and
  pattern recognition}, pages 9185--9193, 2018.

\bibitem{Dvijotham2018}
Krishnamurthy Dvijotham, Robert Stanforth, Sven Gowal, Timothy Mann, and
  Pushmeet Kohli.
\newblock A dual approach to scalable verification of deep networks.
\newblock {\em Conference on Uncertainty in Artificial Intelligence}, 2018.

\bibitem{he2016deep}
Kaiming He, Xiangyu Zhang, Shaoqing Ren, and Jian Sun.
\newblock Deep residual learning for image recognition.
\newblock In {\em Proceedings of the IEEE conference on computer vision and
  pattern recognition}, pages 770--778, 2016.

\bibitem{HenriksenLomuscio20}
P.~Henriksen and A.~Lomuscio.
\newblock Efficient neural network verification via adaptive refinement and
  adversarial search.
\newblock In {\em Proceedings of the 24th European Conference on Artificial
  Intelligence (ECAI20)}, 2020.

\bibitem{HenriksenLomuscio21}
P.~Henriksen and A.~Lomuscio.
\newblock Deepsplit: An efficient splitting method for neural network
  verification via indirect effect analysis.
\newblock In {\em Proceedings of the 30th International Joint Conference on
  Artificial Intelligence (IJCAI21)}, To appear, August 2021.

\bibitem{jaeckle2021generating}
Florian Jaeckle and M~Pawan Kumar.
\newblock Generating adversarial examples with graph neural networks.
\newblock {\em Conference on Uncertainty in Artificial Intelligence}, 2021.

\bibitem{jaeckle2021neural}
Florian Jaeckle, Jingyue Lu, and M~Pawan Kumar.
\newblock Neural network branch-and-bound for neural network verification.
\newblock {\em arXiv preprint arXiv:2107.12855}, 2021.

\bibitem{katz2017reluplex}
Guy Katz, Clark Barrett, David~L Dill, Kyle Julian, and Mykel~J Kochenderfer.
\newblock Reluplex: An efficient smt solver for verifying deep neural networks.
\newblock In {\em International Conference on Computer Aided Verification},
  pages 97--117. Springer, 2017.

\bibitem{katz2019marabou}
Guy Katz, Derek~A Huang, Duligur Ibeling, Kyle Julian, Christopher Lazarus,
  Rachel Lim, Parth Shah, Shantanu Thakoor, Haoze Wu, Aleksandar Zelji{\'c},
  et~al.
\newblock The marabou framework for verification and analysis of deep neural
  networks.
\newblock In {\em International Conference on Computer Aided Verification},
  pages 443--452. Springer, 2019.

\bibitem{KouvarosLomuscio21a}
P.~Kouvaros and A.~Lomuscio.
\newblock Towards scalable complete verification of relu neural networks via
  dependency-based branching.
\newblock In {\em Proceedings of the 30th International Joint Conference on
  Artificial Intelligence ({IJCAI}21)}, To Appear, 2021.

\bibitem{kraska18case}
Tim Kraska, Alex Beutel, Ed~H Chi, Jeffrey Dean, and Neoklis Polyzotis.
\newblock The case for learned index structures.
\newblock In {\em Proceedings of the 2018 International Conference on
  Management of Data}, 2018.

\bibitem{liu2019neuralverification}
Changliu Liu, Tomer Arnon, Christopher Lazarus, and Mykel~J Kochenderfer.
\newblock Neuralverification.jl: Algorithms for verifying deep neural networks.
\newblock In {\em ICLR 2019 Debugging Machine Learning Models Workshop}, 2019.

\bibitem{liu2021algorithms}
Changliu Liu, Tomer Arnon, Christopher Lazarus, Christopher Strong, Clark
  Barrett, and Mykel~J. Kochenderfer.
\newblock Algorithms for verifying deep neural networks.
\newblock {\em Foundations and Trends® in Optimization}, 4(3-4):244--404,
  2021.

\bibitem{Lu2020Neural}
Jingyue Lu and M~Pawan Kumar.
\newblock Neural network branching for neural network verification.
\newblock In {\em International Conference on Learning Representations}, 2020.

\bibitem{madry:17}
Aleksander Madry, Aleksandar Makelov, Ludwig Schmidt, Dimitris Tsipras, and
  Adrian Vladu.
\newblock Towards deep learning models resistant to adversarial attacks.
\newblock In {\em Proc. International Conference on Learning Representations
  (ICLR)}, 2018.

\bibitem{balunovic:20}
Martin~Vechev Mislav~Balunovic.
\newblock Adversarial training and provable defenses: Bridging the gap.
\newblock In {\em Proc. International Conference on Learning Representations
  (ICLR)}, 2020.

\bibitem{mueller2021prima}
Mark~Niklas M{\"u}ller, Gleb Makarchuk, Gagandeep Singh, Markus P{\"u}schel,
  and Martin Vechev.
\newblock Prima: Precise and general neural network certification via
  multi-neuron convex relaxations.
\newblock {\em arXiv preprint arXiv:2103.03638}, 2021.

\bibitem{Paszke2017}
Adam Paszke, Sam Gross, Soumith Chintala, Gregory Chanan, Edward Yang, Zachary
  DeVito, Zeming Lin, Alban Desmaison, Luca Antiga, and Adam Lerer.
\newblock Automatic differentiation in pytorch.
\newblock {\em NIPS Autodiff Workshop}, 2017.

\bibitem{salman2019convex}
Hadi Salman, Greg Yang, Huan Zhang, Cho-Jui Hsieh, and Pengchuan Zhang.
\newblock A convex relaxation barrier to tight robustness verification of
  neural networks.
\newblock {\em Advances in Neural Information Processing Systems},
  32:9835--9846, 2019.

\bibitem{gpupoly}
Fran{\c c}ois Serre, Christoph Müller, Gagandeep Singh, Markus P{\"u}schel,
  and Martin Vechev.
\newblock Scaling polyhedral neural network verification on {GPU}s.
\newblock In {\em Proc. Machine Learning and Systems (MLSys)}, 2021.

\bibitem{shriver-etal:CAV:2021:dnnv}
David Shriver, Sebastian~G. Elbaum, and Matthew~B. Dwyer.
\newblock {DNNV:} {A} framework for deep neural network verification.
\newblock In Alexandra Silva and K.~Rustan~M. Leino, editors, {\em Computer
  Aided Verification - 33rd International Conference, {CAV} 2021, Virtual
  Event, July 20-23, 2021, Proceedings, Part {I}}, volume 12759 of {\em Lecture
  Notes in Computer Science}, pages 137--150. Springer, 2021.

\bibitem{shriver-etal:ICSE:2021:dnnf}
David Shriver, Sebastian~G. Elbaum, and Matthew~B. Dwyer.
\newblock Reducing {DNN} properties to enable falsification with adversarial
  attacks.
\newblock In {\em 43rd {IEEE/ACM} International Conference on Software
  Engineering, {ICSE} 2021, Madrid, Spain, 22-30 May 2021}, pages 275--287.
  {IEEE}, 2021.

\bibitem{singh2019krelu}
Gagandeep Singh, Rupanshu Ganvir, Markus P\"{u}schel, and Martin Vechev.
\newblock Beyond the single neuron convex barrier for neural network
  certification.
\newblock In {\em Advances in Neural Information Processing Systems 32}, pages
  15098--15109. Curran Associates, Inc., 2019.

\bibitem{SinghNIPS:18}
Gagandeep Singh, Timon Gehr, Matthew Mirman, Markus P\"{u}schel, and Martin
  Vechev.
\newblock Fast and effective robustness certification.
\newblock In S.~Bengio, H.~Wallach, H.~Larochelle, K.~Grauman, N.~Cesa-Bianchi,
  and R.~Garnett, editors, {\em Advances in Neural Information Processing
  Systems 31}, pages 10802--10813. Curran Associates, Inc., 2018.

\bibitem{DeepPoly:19}
Gagandeep Singh, Timon Gehr, Markus P\"{u}schel, and Martin Vechev.
\newblock An abstract domain for certifying neural networks.
\newblock {\em Proc. ACM Program. Lang.}, 3(POPL):41:1--41:30, 2019.

\bibitem{singh2019refinement}
Gagandeep Singh, Timon Gehr, Markus Püschel, and Martin Vechev.
\newblock Boosting robustness certification of neural networks.
\newblock In {\em International Conference on Learning Representations (ICLR)}.
  2019.

\bibitem{Singh:17}
Gagandeep Singh, Markus P\"{u}schel, and Martin Vechev.
\newblock Fast polyhedra abstract domain.
\newblock In {\em Proc. Principles of Programming Languages (POPL)}, pages
  46--59, 2017.

\bibitem{tan2019efficientnet}
Mingxing Tan and Quoc Le.
\newblock Efficientnet: Rethinking model scaling for convolutional neural
  networks.
\newblock In {\em International Conference on Machine Learning}, pages
  6105--6114. PMLR, 2019.

\bibitem{Tjeng2019EvaluatingRO}
Vincent Tjeng, Kai~Y. Xiao, and Russ Tedrake.
\newblock Evaluating robustness of neural networks with mixed integer
  programming.
\newblock In {\em ICLR}, 2019.

\bibitem{tran2020cav}
Hoang-Dung Tran, Stanley Bak, Weiming Xiang, and Taylor~T. Johnson.
\newblock Verification of deep convolutional neural networks using imagestars.
\newblock In {\em 32nd International Conference on Computer-Aided Verification
  (CAV)}. Springer, July 2020.

\bibitem{tran2019formalise}
Hoang-Dung Tran, Patrick Musau, Diego~Manzanas Lopez, Xiaodong Yang, Luan~Viet
  Nguyen, Weiming Xiang, and Taylor~T. Johnson.
\newblock Parallelizable reachability analysis algorithms for feed-forward
  neural networks.
\newblock In {\em Proceedings of the 7th International Workshop on Formal
  Methods in Software Engineering (FormaliSE'19)}, FormaliSE '19, pages 31--40,
  Piscataway, NJ, USA, May 2019. IEEE Press.

\bibitem{tran2019fm}
Hoang-Dung Tran, Patrick Musau, Diego~Manzanas Lopez, Xiaodong Yang, Luan~Viet
  Nguyen, Weiming Xiang, and Taylor~T. Johnson.
\newblock Star-based reachability analysis for deep neural networks.
\newblock In {\em 23rd International Symposium on Formal Methods (FM'19)}.
  Springer International Publishing, October 2019.

\bibitem{tran2020cav_tool}
Hoang-Dung Tran, Xiaodong Yang, Diego~Manzanas Lopez, Patrick Musau, Luan~Viet
  Nguyen, Weiming Xiang, Stanley Bak, and Taylor~T. Johnson.
\newblock {NNV}: The neural network verification tool for deep neural networks
  and learning-enabled cyber-physical systems.
\newblock In {\em 32nd International Conference on Computer-Aided Verification
  (CAV)}, July 2020.

\bibitem{vincent2021reachable}
Joseph~A. Vincent and Mac Schwager.
\newblock Reachable polyhedral marching (rpm): A safety verification algorithm
  for robotic systems with deep neural network components, 2021.

\bibitem{vnn-comp}
VNN-COMP.
\newblock International verification of neural networks competition
  ({VNN-COMP}).
\newblock {\em Verification of Neural Networks workshop at the International
  Conference on Computer-Aided Verification}, 2020.

\bibitem{wang2021betacrown}
Shiqi Wang, Huan Zhang, Kaidi Xu, Xue Lin, Suman Jana, Cho-Jui Hsieh, and Zico
  Kolter.
\newblock {Beta-CROWN}: Efficient bound propagation with per-neuron split
  constraints for complete and incomplete neural network verification.
\newblock {\em arXiv preprint arXiv:2103.06624}, 2021.

\bibitem{Wong2018}
Eric Wong and Zico Kolter.
\newblock Provable defenses against adversarial examples via the convex outer
  adversarial polytope.
\newblock 2018.

\bibitem{wu2020parallelization}
Haoze Wu, Alex Ozdemir, Aleksandar Zeljic, Kyle Julian, Ahmed Irfan, Divya
  Gopinath, Sadjad Fouladi, Guy Katz, Corina Pasareanu, and Clark Barrett.
\newblock Parallelization techniques for verifying neural networks.
\newblock In {\em \# PLACEHOLDER\_PARENT\_METADATA\_VALUE\#}, volume~1, pages
  128--137. TU Wien Academic Press, 2020.

\bibitem{xiang2018tnnls}
W.~{Xiang}, H.~{Tran}, and T.~T. {Johnson}.
\newblock Output reachable set estimation and verification for multilayer
  neural networks.
\newblock {\em IEEE Transactions on Neural Networks and Learning Systems},
  29(11):5777--5783, 2018.

\bibitem{xu2020automatic}
Kaidi Xu, Zhouxing Shi, Huan Zhang, Yihan Wang, Kai-Wei Chang, Minlie Huang,
  Bhavya Kailkhura, Xue Lin, and Cho-Jui Hsieh.
\newblock Automatic perturbation analysis for scalable certified robustness and
  beyond.
\newblock {\em Advances in Neural Information Processing Systems}, 33, 2020.

\bibitem{xu2021fast}
Kaidi Xu, Huan Zhang, Shiqi Wang, Yihan Wang, Suman Jana, Xue Lin, and Cho-Jui
  Hsieh.
\newblock {Fast and Complete}: Enabling complete neural network verification
  with rapid and massively parallel incomplete verifiers.
\newblock In {\em International Conference on Learning Representations}, 2021.

\bibitem{zhang2018efficient}
Huan Zhang, Tsui-Wei Weng, Pin-Yu Chen, Cho-Jui Hsieh, and Luca Daniel.
\newblock Efficient neural network robustness certification with general
  activation functions.
\newblock {\em Advances in Neural Information Processing Systems},
  31:4939--4948, 2018.

\end{thebibliography}
